\pdfoutput=1

\documentclass[futureinternet,article,accept,pdftex,moreauthors]{Definitions/mdpi} 

\firstpage{1} 
\pubvolume{1}
\issuenum{1}
\articlenumber{0}
\pubyear{2024}
\copyrightyear{2024}
\externaleditor{Academic Editor: Firstname\linebreak Lastname}
\datereceived{29 January 2024} 
\daterevised{19 February 2024} 
\dateaccepted{21 February 2024} 
\datepublished{ } 
\hreflink{https://doi.org/} 

\Title{Analyzing GPU Performance in Virtualized Environments: A~Case Study}

\TitleCitation{Analyzing GPU Performance in Virtualized Environments: A Case Study}

\Author{Adel Belkhiri *\orcidA{} 
 and Michel Dagenais}

\AuthorNames{Adel Belkhiri and Michel Dagenais}

\AuthorCitation{Belkhiri, A.; Dagenais, M.}

\address[1] {Department 
 of Computer and Software Engineering, École Polytechnique de Montréal, {Montréal, QC H3T 1J4, Canada;} 
 {michel.dagenais@polymtl.ca} 
}

\corres{\hangafter=1 \hangindent=1.05em \hspace{-0.82em}Correspondence: adel.belkhiri@polymtl.ca}

\abstract{The graphics processing unit (GPU) plays a crucial role in boosting application performance and enhancing computational tasks. Thanks to its parallel architecture and energy efficiency, the GPU has become essential in many computing scenarios. On the other hand, the advent of GPU virtualization has been a significant breakthrough, as it provides scalable and adaptable GPU resources for virtual machines. However, this technology faces challenges in debugging and analyzing the performance of GPU-accelerated applications. Most current performance tools do not support virtual GPUs (vGPUs), highlighting the need for more advanced tools. Thus, this article introduces a novel performance analysis tool that is designed for systems using vGPUs. Our tool is compatible with the Intel GVT-g virtualization solution, although its underlying principles can apply to many vGPU-based systems. Our tool uses software tracing techniques to gather detailed runtime data and generate relevant performance metrics. It also offers many synchronized graphical views, which gives practitioners deep insights into GVT-g operations and helps them identify potential performance bottlenecks in vGPU-enabled virtual machines.
}


\begin{document}


\section{Introduction}
Accelerators are specialized processors that have been developed and integrated into computer systems in response to the growing demand for high-performance computing. These accelerators aim to assist the central processing unit (CPU) in executing certain types of computations. Several studies have demonstrated that offloading specific tasks to accelerators can considerably boost the overall system performance \cite{Hong2017}. Hence, the use of accelerators has become a typical approach to handling the high computational demands in various industries and research fields. This has motivated hardware manufacturers to develop a variety of specialized accelerators, including accelerated processing units (APUs), floating-point units (FPUs), digital signal processing units (DSPs), network processing units (NPUs), and graphics processing units (GPUs).

The GPU, which is considered one of the most pervasive accelerators, was initially devised for graphics rendering and image processing. However, in the last few years, its computational power has been increasingly harnessed for parallel number crunching. Nowadays, GPU-accelerated applications are present in many domains that are often unrelated to graphics, such as deep learning, financial analytics, and general-purpose scientific calculations. On the other hand, virtualization plays a fundamental role, as it enables many modern computing concepts. Its primary function is facilitating the sharing and multiplexing of physical resources among different applications. Particularly, virtualization significantly enhances GPU utilization by enabling more efficient allocation of its computational power. In general, the main advantage of virtualizing computing resources lies in reducing energy consumption in data centers, which, in turn, contributes to reducing operational costs. Hence, a variety of new applications that leverage the capabilities of virtual GPUs (vGPUs) have emerged. Examples of such applications that benefit from vGPU acceleration are deep learning, virtual desktop infrastructure (VDI), and artificial intelligence applications.

It is worth noting that virtualizing GPUs presents more complex challenges than the virtualization of CPUs and most I/O devices, as the latter rely on well-established technologies. These challenges stem from several key obstacles. First, there is significant architectural diversity in hardware among GPU brands, complicating the development of a universal virtualization solution. Second, using closed-source drivers from popular GPU brands, such as NVIDIA, poses a significant challenge for third-party developers in developing virtualization technology for these devices. Third, most GPU designs lack inherent sharing mechanisms, leading to a GPU process gaining exclusive access to its resources, thereby blocking other processes from preemption. In addition, several studies have shown that the overhead involved in process preemption is substantially higher in GPUs than in CPUs \cite{Ji2022Compiler-Directed, Hong2017Bis}. This increased overhead is essentially due to the higher number of cores and context states in GPUs. It is important to note that some recent GPUs, such as the NVIDIA Pascal GPU \cite{NvidiaPascal2016}, include support for preemption at the kernel, thread, and instruction levels to mitigate these challenges.

Before developing advanced GPU virtualization technologies, practitioners used the passthrough technique to enable VMs to access physical GPU (pGPU) resources directly. This approach, however, has limitations, such as the inability to share a GPU among multiple VMs and the lack of support for live migration. Major GPU manufacturers such as NVIDIA, AMD, and Intel have introduced their brand-specific virtualization solutions to address these issues. These include NVIDIA GRID \cite{NvidiaGrid}, AMD MxGPU \cite{MxGPU}, and Intel's Graphics Virtualization Technology---Grid generation (GVT-g) \cite{KunTian2014}. The first two virtualization solutions are based on hardware virtualization capabilities, whereas the third one, Intel's GVT-g, provides a software-based solution for full GPU virtualization. GVT-g is open source and has been integrated into the Linux mainline kernel, which makes it a desirable option due to its accessibility and potential for broader integration.

On the other hand, performance analysis tools for GPUs are important for debugging performance issues in GPU-accelerated applications \cite{Aceto2013}. These tools help understand how the GPU resources are allocated and consumed, and they facilitate the diagnosis of potential performance bottlenecks. They are particularly crucial in virtualized environments, where resource sharing in vGPUs and its impact on performance need to be better understood. However, developing practical tools for monitoring and debugging vGPUs remains challenging. This is because virtualized environments often present many layers, encompassing hardware, middleware, and host and guest operating systems, which increase the isolation and abstraction of GPU resources, making it difficult to pinpoint the causes of \mbox{performance issues.}

The landscape of GPU performance analysis features diverse tools, with academic contributions tailored to specific GPU programming models or diagnosing particular GPU-related performance issues. For instance, the works in 
 \cite{Couturier2015,Margheritta2019} leveraged library interposition to capture runtime events from OpenCL- and HSA-based programs, respectively. While they enabled the analysis of GPU kernel execution and CPU--GPU interaction, they presented high overhead and faced challenges for broader analysis and portability. On the other hand, vendor-provided options, such as vTune Profiler \cite{vTuneAmplifier2018} and Nsight Systems \cite{NvidiaNsight2018}, leverage dynamic instrumentation and hardware counters to gather performance data and unveil application behavior in various aspects. These tools offer interesting analyses covering kernel execution, memory access patterns, and GPU API call paths. Nevertheless, they are often limited in terms of openness and cross-architecture applicability. In summary, while each of these tools has its strengths and weaknesses, a common shortfall is their lack of support for vGPUs---with the exception of Nsight Systems, which, however, is proprietary and closed-source software.

This paper presents a novel performance analysis framework dedicated to  GPUs virtualized with the GVT-g technology. Our framework uses  host-based tracing to gather performance data efficiently and with minimal overhead. Tracing is a proven technique for collecting detailed performance data from complex systems. A key advantage of our approach is that it does not necessitate the installation of any agents or tools within the VMs, as tracing is confined to the host operating system's kernel space. The benefits of this methodology include ease of deployment, preservation of VM owner privacy, and cost-effectiveness. This study's contributions are as follows:

\begin{enumerate}
\item Efficient instrumentation of KVMGT, the open-source implementation of GVT-g for KVM \cite{Song2014}, and integrated modules within the Linux Trace Toolkit Next Generation (LTTng) \cite{LTTngProject2018} to collect performance data from both the virtualization solution and the Intel GPU driver.

\item Development of a unified, stateful model capable of filtering and organizing the collected trace data. This model simplifies data processing and supports our analysis in computing many performance metrics.

\item Implementation of various analyses within Trace Compass \cite{TraceCompass}, an open-source performance analyzer, and a series of synchronized graphical views. These views aim to assist practitioners in understanding GVT-g's internal mechanisms and in diagnosing many performance issues related to vGPU usage.
\end{enumerate} 

The remainder of this paper is organized as follows: In Section \ref{background}, we conduct a comprehensive literature review of the methodologies and techniques applied in GPU virtualization. Section \ref{related-work} introduces prominent GPU profiling and performance analysis tools. Section \ref{Architecture-of-Our-Framework} details the design of our proposed framework and outlines several pertinent performance metrics. Next, Section \ref{Use-Cases} presents three practical scenarios that demonstrate the effectiveness of our framework. Following this, Section \ref{Tracing-Cost-Analysis} assesses and discusses the overhead associated with using our framework. Finally, we summarize our findings and contributions in Section \ref{Conclusion}.

\section{Background} \label{background}
In this section, we delve into the methodologies and technologies that form the foundation of GPU virtualization. Understanding these virtualization technologies is crucial for developing performance analysis solutions, as they directly impact the allocation and management of host resources. Despite challenges, such as proprietary GPU drivers and non-standardized GPU designs, a number of virtualization techniques have been developed~\cite{Hong2017}. We categorize these techniques into four primary categories: PCI passthrough, single-root input/output virtualization (SR-IOV), para- and full virtualization, and \mbox{API remoting.}

\subsection{PCI Passthrough}
PCI passthrough utilizes hardware virtualization technologies---specifically, Intel VT-d~\cite{Abramson2006} and AMD-Vi \cite{Doorn2006}---to allow VMs to access the pGPU of the host directly. While this method offers near-native GPU performance, it limits one GPU to a single VM, leading to potential resource underutilization. Google Cloud is an example of a cloud service employing this \mbox{technology \cite{GoogleCloudGPU2018}.}

\subsection{SR-IOV}
SR-IOV addresses the resource-sharing limitation of PCI passthrough. Partitioning a single physical device's hardware functions into physical functions (PFs) and virtual functions (VFs) allows multiple VMs to share a single device. Supported by major hypervisors such as KVM and Xen \cite{Zhang2016, Zhang2017}, its implementations include NVIDIA GRID \cite{NvidiaGrid} and AMD MxGPU \cite{MxGPU}. SR-IOV offers balanced performance, equitable pGPU distribution, and enhanced security via vGPU isolation.

\subsection{API Remoting}
\textls[-15]{API remoting virtualizes GPUs by intercepting and processing API calls in the guest OS. While effective, it requires frequent updates to wrapper libraries to stay aligned with evolving GPU frameworks \cite{GViM2009, Pegasus2011, rCUDA2010, VADI2016, vCUDA2012}. This approach is notable for its high-level \mbox{virtualization capabilities.}}

\subsection{Para- and Full Virtualization}
GPUs in this category are virtualized at the driver and hypervisor levels. Paravirtualization employs customized GPU drivers in the guest OS, whereas full virtualization uses standard drivers. A notable example is Intel GVT-g \cite{KunTian2014}, a technology that utilizes mediated passthrough (MPT) for efficient and isolated vGPU access. This method supports advanced features such as live migration \cite{Ma2018} and enhances vGPU density \cite{YaozuDong2015, MochieXue2016}, contributing to its comprehensive utility in virtualized environments.

Although not widely adopted in cloud environments due to competition from proprietary solutions, GVT-g holds significant value for its distinctive technical advantages. It seamlessly integrates with the Linux kernel as an open-source software-based solution, enabling community customization and broader applicability compared to closed-source alternatives. Furthermore, its lightweight design minimizes performance overhead and hardware dependence, making it a cost-effective option for diverse computing scenarios. 


\section{Related Work} \label{related-work}
As the complexity of GPU-accelerated applications continues to grow, the need for effective performance analysis methods becomes increasingly critical, particularly in vGPU-based systems. Our study of existing GPU performance analysis tools shows that they offer different levels of analysis, and they are mostly dedicated to specific GPU architectures. High-end production-quality tools such as vTune Profiler \cite{vTuneAmplifier2018} and Nsight systems \cite{NvidiaNsight2018} are notable for their comprehensive approach to analyzing GPU-accelerated applications. They provide a holistic understanding of the application runtime behavior, particularly unveiling the interaction between CPU and GPU, which is crucial for effective optimization. In contrast, tools proposed in academic research are often tailored to specific GPU programming models or  addressing particular GPU-related performance issues.

Many vendors offer dedicated software for profiling GPU applications, such as NVIDIA Nsight Systems, Intel vTune Profiler, and AMD Radeon GPU Profiler \cite{AMDRadeonProfiler2024}. These tools leverage various techniques such as binary instrumentation, hardware counters, and API hooking to gather detailed performance events. Despite providing rich insights into kernel execution, CPU--GPU interaction, memory access patterns, and GPU API call paths, these tools are often limited in terms of openness, flexibility, and cross-architecture applicability. In addition to proprietary offerings, the GPU performance analysis ecosystem encompasses feature-rich open-source tools. For example, HPCToolkit \cite{Gupta2020, Cherian2021} and TAU \cite{ParaToolsTAU2018} are two versatile tools tailored for analyzing heterogeneous systems' performance. These tools offer valuable diagnostic capabilities for pinpointing GPU bottlenecks and determining their root causes. For instance, through call path profiling, they provide insights for kernel execution and enable the identification of hotspots in the program's code.

Aside from the established profilers, academic research also presents many innovative tools for the diagnosis of performance issues  in GPU-accelerated applications. Zhou et al. proposed GVProf \cite{GVPROF2020}, a value-aware profiler for identifying redundant memory accesses in GPU-accelerated applications. Their follow-up work \cite{Falsafi2022} focused on improving the detection of value-related patterns (e.g., redundant values, duplicate writes, and single-valued data). The main objective of their work was to identify diverse performance bottlenecks and provide suggestions for code optimization. GPA (GPU Performance Advisor) \cite{Zhou2021} is a diagnostic tool that leverages instruction sampling and data flow analysis to pinpoint inefficiencies in the application code. DrGPU \cite{DrGPU2023} uses a top-down profiling approach to quantify and decompose stall cycles using hardware performance counters. Based on the stall analysis, it identifies inefficient software--hardware interactions and their root causes, thus helping make informed optimization decisions. CUDAAdvisor \cite{CUDAAdvisor2018}, built on top of LLVM, instrumentalizes application code on both the host and device sides. It conducts code- and data-centric profiling to identify performance bottlenecks arising from competition for cache resources and memory and control flow divergence. The main disadvantages of these tools lie in their considerable overhead and exclusive applicability to NVIDIA GPUs. On the other hand, several profiling tools leverage library interposition and userspace tracing to capture runtime events, enabling the correlation of CPU and GPU activities. For example, CLUST \cite{Couturier2015} and LTTng-HSA \cite{Margheritta2019} employ these techniques to profile OpenCL- and HSA-based applications, respectively. However, a substantial drawback of these tools is their tight coupling with specific GPU programming frameworks, which limits their capability to provide a system-wide analysis.

To sum up, the current landscape of GPU performance analysis tools is diverse, with each tool differing in openness, targeted GPU architectures, and depth of analysis. A significant limitation among these tools, with the notable exception of those offered by NVIDIA, is their lack of support for vGPUs (Table \ref{related-work-comparison}). This gap is particularly significant considering the growing importance of GPUs in virtualized environments. Our research aims to address this oversight by proposing a method for analyzing the performance of GPUs virtualized via GVT-g. We detail this method and the framework that it implements in the following section.

\begin{table}[H]
\caption{A comparative analysis of a few state-of-the-art tools alongside our framework.\label{related-work-comparison}}
	\begin{adjustwidth}{-\extralength}{0cm}
  \small
		\begin{tabularx}{\fulllength}{LLLLCl}
			\toprule
			\textbf{Tool} & \textbf{GPU Arch.} & \textbf{GPU APIs} & \textbf{Analysis Levels} & \textbf{vGPU Support} & \textbf{Overhead} \\

\midrule
   \multirow[m]{1}{*}{ValueExpert \cite{Falsafi2022}} & NVIDIA & CUDA & Thread & No & Moderate\\   

\midrule
   \multirow[m]{1}{*}{GPA \cite{Zhou2021}} & NVIDIA & CUDA & Thread & No & High\\ 

\midrule
   \multirow[m]{1}{*}{CLUST \cite{Couturier2015}} & - & OpenCL & Thread & No & Moderate\\ 

\midrule
   \multirow[m]{1}{*}{LTTng-HSA \cite{Margheritta2019}} & - & HSA & Thread & No & Moderate\\  

\midrule
   \multirow[m]{1}{*}{DrGPU \cite{DrGPU2023}} & NVIDIA & CUDA & Thread, instruction & No & High\\  

\midrule
   \multirow[m]{1}{*}{GVProf \cite{GVPROF2020}} & NVIDIA & CUDA & Thread, instruction & No & Moderate\\

\midrule
   \multirow[m]{1}{*}{CUDAAdvisor \cite{CUDAAdvisor2018}} & NVIDIA	& CUDA & Thread, instruction & No & High\\
   
\midrule
   \multirow[m]{1}{*}{Our framework 
} & Intel & - & System-wide, thread & Yes & Low\\    
			\bottomrule
		\end{tabularx}
	\end{adjustwidth}
\end{table}

\section{Proposed Solution} \label{Architecture-of-Our-Framework}
\begin{figure}[H]
\includegraphics[width=13cm]{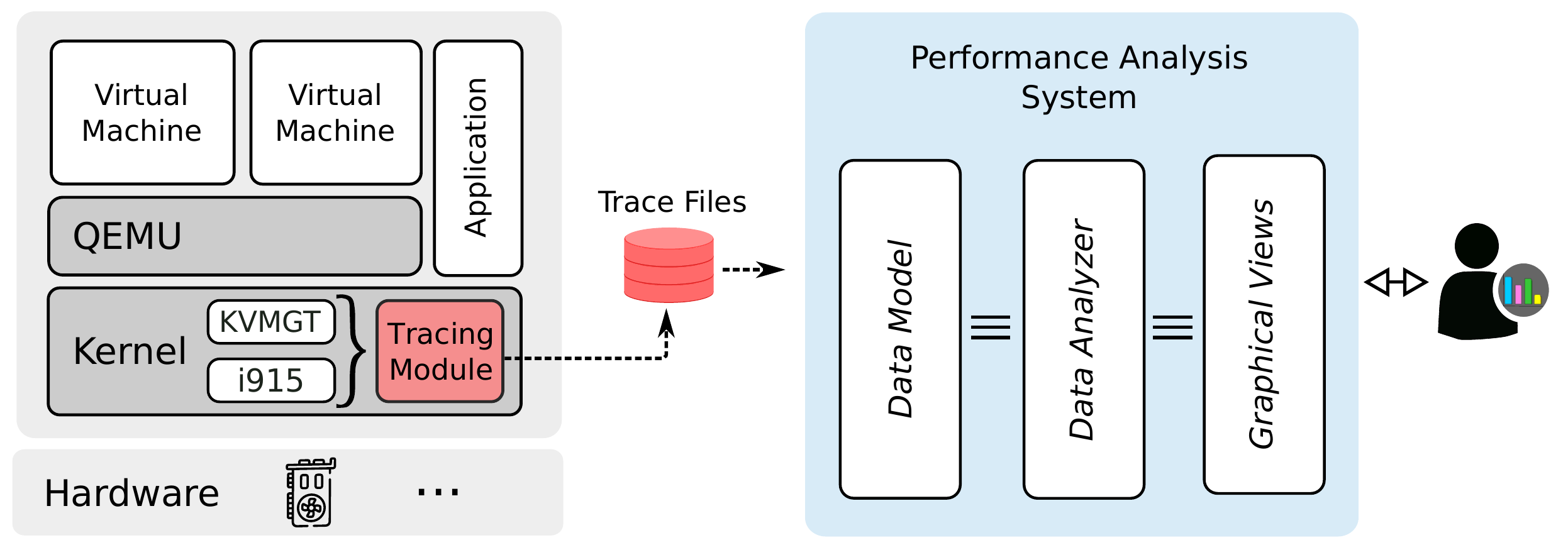}
\caption{Architecture of the proposed framework.}
\label{fig:architecture-framework}
\end{figure}

This paper presents a new performance analysis framework for vGPUs managed with GVT-g. The architecture of this framework is depicted in Figure \ref{fig:architecture-framework}. It comprises three main subsystems: a data collection subsystem, an analysis subsystem, and a visualization subsystem. The first subsystem is in charge of collecting meaningful performance data from GVT-g and the Intel GPU driver. The second subsystem processes the collected data and derives relevant performance metrics. The visualization subsystem consists of several graphical views that display the results of the analysis conducted by the second subsystem. We implemented the visualization and analysis subsystems as extensions of Trace Compass~\cite{TraceCompass}, an open-source trace analyzer. This tool can analyze a massive amount of tracing data to enable the user to diagnose a variety of performance bottlenecks. Moreover, Trace Compass supports interactive graphical interfaces, pre-built analyses, event filtering, and trace synchronization. We present the design details of those subsystems in the following subsections.

In addition, we used KVM \cite{KVM2018}  to build our virtualization environment since it is the default hypervisor of Linux. The role of KVM is to enable VMs to leverage the hardware virtualization capabilities of the host machine (e.g., VT-x and AMD-V on Intel and AMD machines, respectively). In Linux, KVM is implemented as three kernel modules: kvm.ko, kvm-intel.ko, and kvm-amd.ko. At the user-space level, we used QEMU \cite{QEMU2018} to run the guest OSs of our VMs. 

\subsection{Data Collection} \label{data-collection}
Logging and tracing are fundamental techniques for collecting runtime data from software. Logging involves recording data in one or more log files and detailing application failures, misbehavior, and the status of its ongoing operations. These high-level human-readable data help in swift bug troubleshooting. In contrast, tracing captures low-level details of software execution, focusing on diagnosing complex functional issues and identifying performance bottlenecks. The resultant data tend to be extensive and detailed. Unlike logging, tracing requires a specialized tool known as a tracer. A multitude of tracers are available for nearly all modern operating systems. For example, in Linux systems, notable tracers include Ftrace \cite{FtraceRedHat2018}, Perf \cite{Perf2015}, and LTTng.

Our framework relies on the tracing technique for gathering low-level data for our analysis. It uses LTTng as a tracer because of its versatility and minimal overhead. It is worth noting that we have confined our tracing scope to the host machine to minimize intrusiveness. In addition, as GVT-g is a kernel module (named \textit{kvmgt.ko} in Linux systems), our focus is restricted to kernel space tracing. This involves instrumenting the GVT-g module's source code to introduce new tracepoints and using a selection of existing tracepoints in i915, the host GPU driver. Table \ref{tab:kern-tracepoints-descr} details a subset of these tracepoints, which are essential for our analysis.

\begin{table}[H]%
\caption{A subset of the kernel events required for our analysis.\label{tab:kern-tracepoints-descr}}
\begin{tabularx}{\linewidth}{>{\raggedright\arraybackslash}p{3.5cm}X}
\toprule
\textbf{Tracepoint} & \textbf{Description} \\
\midrule
gvt\_workload\_queue & Indicates the addition of a GPU request to the vGPU queue after submission by a VM process. \\
gvt\_workload\_submit & Marks the forwarding of a request from GVT-g threads to the host graphics driver (i915). \\
gvt\_workload\_complete & Fires upon completion and removal of a GPU request from GVT-g data structures. \\
gvt\_sched\_switch & Reports the remaining time slice for each vGPU upon scheduling out. \\
i915\_gem\_request\_add & Indicates the queuing of a GPU request by the host GPU driver. \\
i915\_gem\_request\_submit & Fires when a request's dependencies are resolved and it is ready for execution. \\
i915\_gem\_request\_in & Indicates the sending of a GPU request to the hardware for execution. \\
i915\_gem\_request\_out & Marks the removal of a request from the host GPU driver's data structures. \\
i915\_intel\_engine\_notify & Indicates the completion of a GPU request by one of the GPU engines. \\
i915\_gem\_object\_create & Triggered when a new GEM object is created, indicating the allocation of a new chunk of memory within the GPU's memory space. \\
i915\_gem\_object\_destroy & Occurs when a GEM object is destroyed, signaling the release of the memory associated with that object. \\
\bottomrule
\end{tabularx}
\end{table}

\subsection{Data Analysis \label{Performance-Analysis-Subsystem}}
The data generated by tracers are inherently low-level and complex, making any manual analysis very complex. The trace events are semantically interconnected and can be fully understood only within their specific context. Consequently, our principal objective is to develop an automated analysis system that improves the interpretability of the collected data. This system aims to assist practitioners in rapidly identifying performance issues arising from utilizing GPU resources in virtualized environments.

\subsubsection{Data Model}
As it is designed for offline analysis, our framework requires a robust data model to organize the data collected from the traced kernel subsystems efficiently. An efficient model is indeed essential for generating metrics and graphical representations effectively. Thus, it is necessary to have a thorough understanding of the GPU request's lifecycle---from when a VM process initiates a request to the point where the hardware executes it and returns a response. Such a comprehensive insight is paramount for diagnosing intricate performance anomalies effectively. 

\begin{figure}[H]
\begin{minipage}[l]{0.65\linewidth}
\includegraphics[width=64mm]{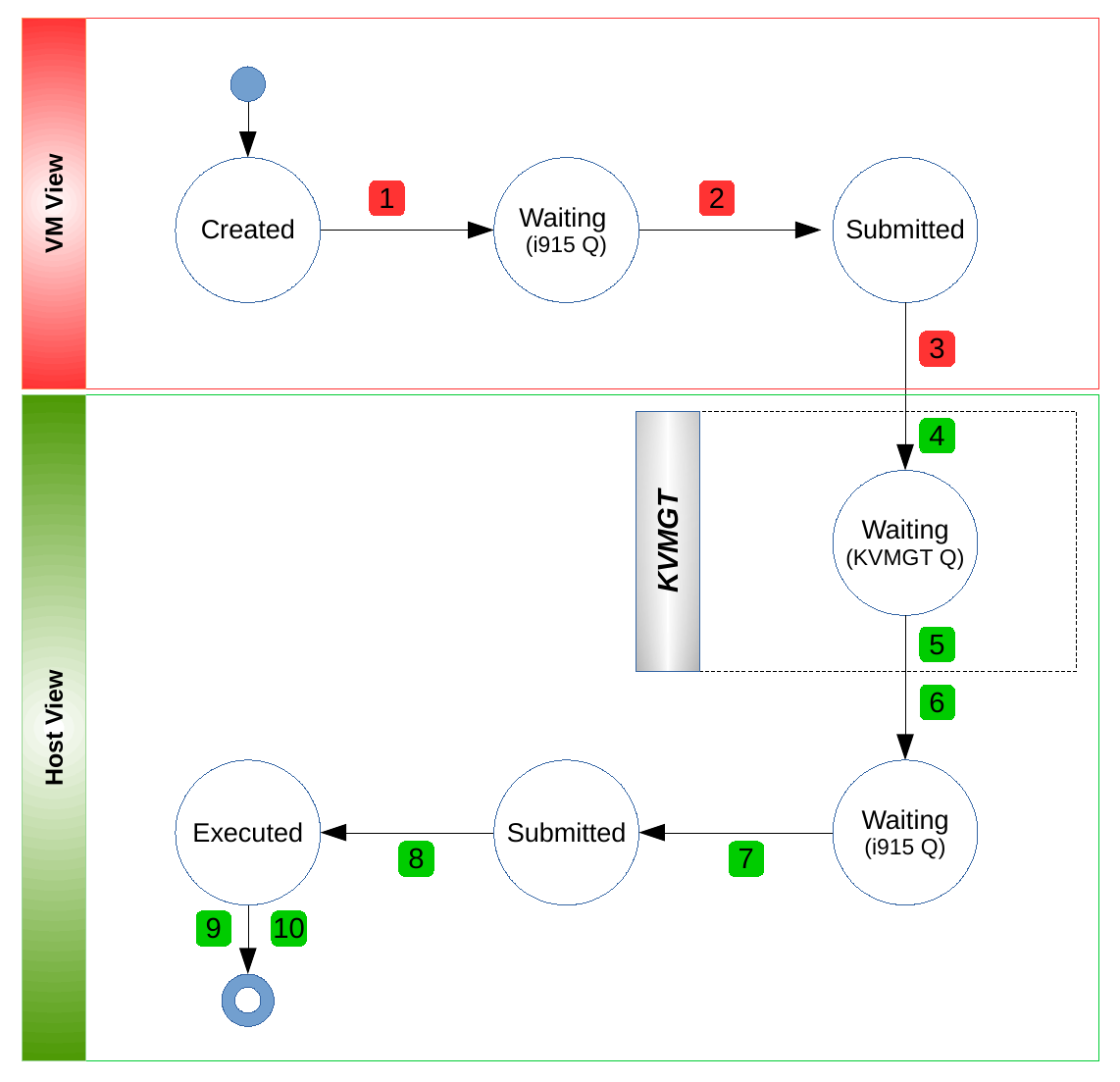}
\end{minipage}
\hspace{-1.5cm}
\begin{minipage}[l]{0.32\linewidth}
\scriptsize
\begin{tabular}{l l}
    \toprule
    \textbf{\#} & \textbf{Event} \\
    \midrule
    1  & \texttt{gem\_request\_add} \\
    2  & \texttt{i915\_gem\_request\_submit} \\     
    3  & \texttt{i915\_gem\_request\_in} \\              
    4  & \texttt{gvt\_workload\_queue} \\               
    5  & \texttt{gvt\_workload\_submit} \\ 
    6  & \texttt{gem\_request\_add} \\
    7  & \texttt{i915\_gem\_request\_submit} \\         
    8  & \texttt{i915\_gem\_request\_in} \\             
    9  & \texttt{intel\_engine\_notify} \\ 
    10 & \texttt{i915\_gem\_request\_out} \\
    \bottomrule
    \end{tabular}
    \label{table:student}
\end{minipage}
\captionof{figure}{States of a GPU request along with the corresponding tracing events.} 
\label{fig:states-request} 
\end{figure}

Figure \ref{fig:states-request} illustrates the various states that a GPU request undergoes throughout its lifecycle in the system. Initially, a process within a GPU-accelerated VM generates a request for GPGPU or graphics processing operations. This request is first received by the VM's GPU driver, which places it in a waiting queue (1). 
 As GPU requests often depend on others, these dependencies must be resolved before the request attains the ``Submitted'' state (2). Once all preceding requests in the queue have been executed, the guest OS's GPU driver removes the request from its waiting queue and forwards it to the vGPU for execution (3). Subsequently, GVT-g intercepts the request and queues it in the vGPU's waiting queue associated with the VM (4). When the vGPU's scheduler allocates time for this vGPU, the request is passed to the host GPU driver (5), which then places it in another waiting queue~(6). The request awaits its turn for execution, pending the resolution of its dependencies~(7) and completing all prior queued requests. Once these conditions are met, the host GPU driver dispatches the request to the hardware for execution (8). Upon completing the request, GVT-g receives a notification (9) and informs the guest OS via a virtual interrupt.
\vspace{-5pt}

\begin{figure}[H]
\includegraphics[width=8cm]{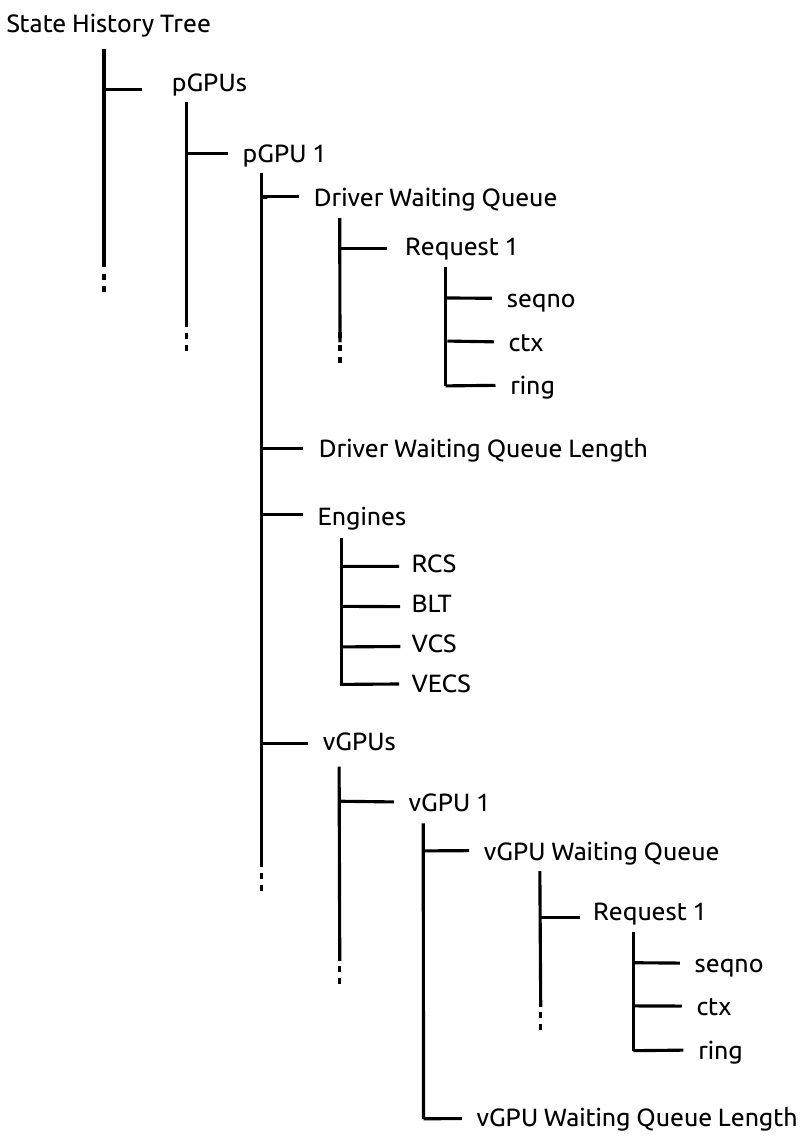}
\caption{Excerpt of the tree-based data structure upon which our analyses are based.}
\label{fig:state-system}
\end{figure}

Modeling the collected performance data is mandatory for computing relevant performance metrics, such as the average wait and execution times of requests per vGPU. To this end, we used Trace Compass \cite{TraceCompass}, a powerful trace analyzer that integrates several interesting pre-built performance analyses. These analyses save the data representing the states of the system to be monitored in a disk-based data structure called a state history tree (SHT) \cite{Montplaisir2013}. We store the states of our system in an attribute--tree data structure (Figure \ref{fig:state-system}). This data structure exhibits sets of hierarchical attributes, which are updated each time one of the events described in Section \ref{data-collection} is handled. Thus, according to our model, one or many vGPUs can be attached to one pGPU. Both a pGPU and a vGPU each have their own driver waiting queue. A GPU request is identified by its sequence number (\textit{seqno}), its context (\textit{ctx}), and the engine to which it should be sent for execution (\textit{ring}).
\vspace{-5pt}

\subsubsection{Performance Metrics}
A performance metric is a quantitative measure of a system's performance in a specific aspect. Mainly, performance metrics help assess the status of a system and pinpoint potential performance problems. This section introduces several performance metrics that our framework calculates.

\paragraph{GPU Utilization 
} This metric quantifies the occupancy levels of both the pGPU and the vGPU. It is expressed as the percentage of the time during which the GPU is actively processing requests. Considering that the GPU operates multiple engines to process requests in different queues concurrently, we calculate this metric on a per-engine basis. The formula below uses $t_{s}$ and $t_{e}$ as markers, representing the start and end timestamps, respectively, for the execution of an individual request denoted by i. Thus, the GPU utilization metric U is calculated as the sum of the execution durations of all requests tracked during T, the observation period. This sum is then divided by T to obtain the utilization percentage. 

\begin{equation*}
 U = \sum_{i}^{}\frac{t_{e_{i}}-t_{s_{i}}}{T}\times100
\end{equation*} 

\paragraph{Waiting Queue Length} This metric indicates the number of requests awaiting execution or processing. It is used by practitioners to identify bottlenecks or performance degradation, especially when examined alongside other metrics. The waiting queue length metric varies based on factors such as the frequency of request issuance, the size of the requests, and the hardware's execution rate. It functions as a counter that increments with each request added to the queue and decrements when a request is removed.

\paragraph{Average Wait Time} As previously noted, the hardware does not execute GPU requests immediately; they spend some time in various queues, awaiting processing by the pGPU. This delay, or waiting time, starts when a request is queued in the waiting queue of the guest GPU driver and ends when it is dispatched for execution. It is indeed a key factor in evaluating the performance efficiency of the vGPU, as it directly impacts the total time required to process GPU requests. In addition, analyzing the waiting time can provide some insights into potential performance bottlenecks or inefficiencies in the GPU request handling process. Our framework estimates the average wait time (WT) of GPU requests using the formula below. It calculates it by summing the waiting durations (W) of individual requests observed within a specific period and then dividing this sum by n, the total number of requests. 
\begin{equation*}
WT=\sum_{i}\frac{W_{i}}{n}
\end{equation*}

\begin{figure}[H]
\includegraphics[width=13cm]{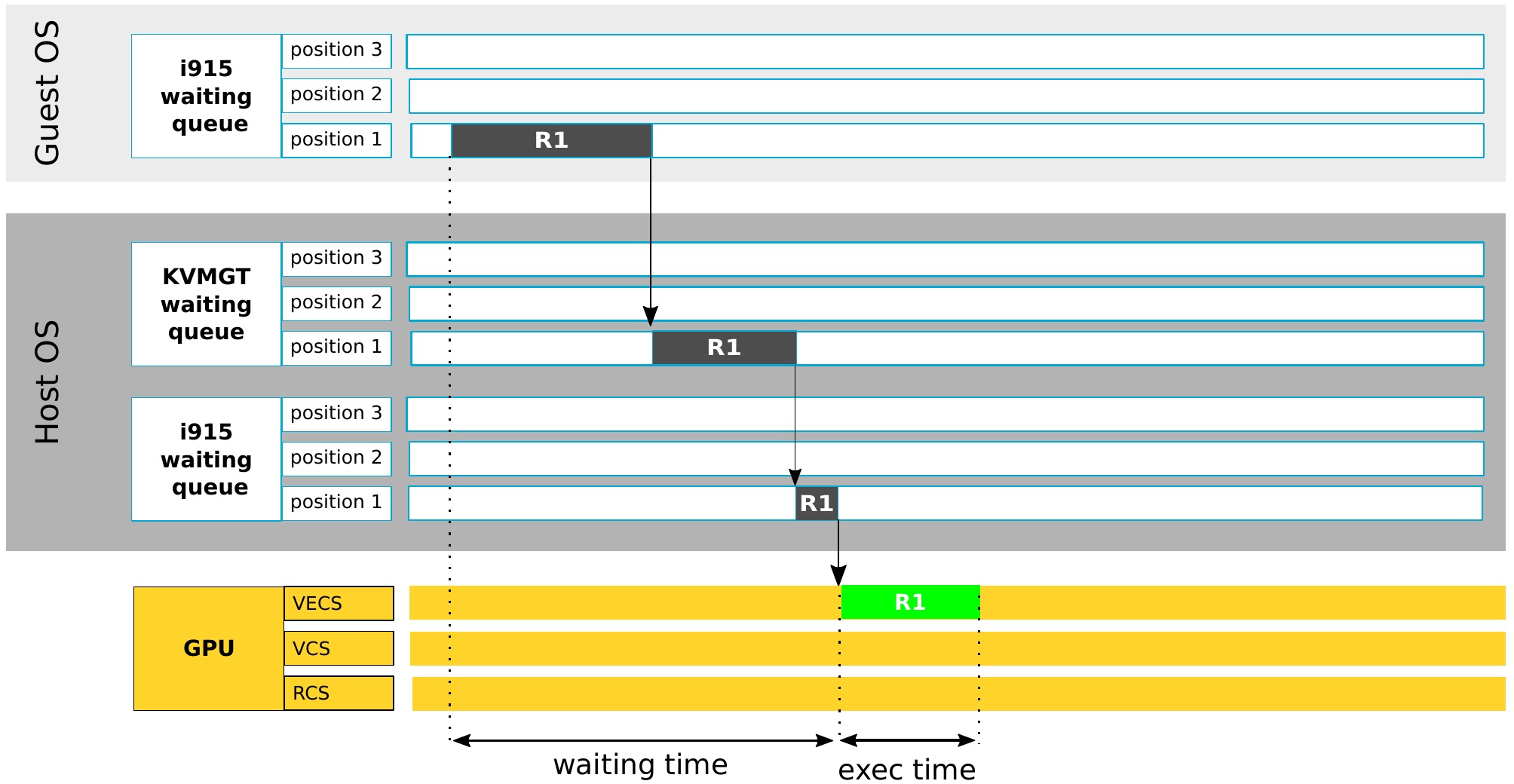}
\caption{The queuing and processing phases of GPU requests.}
\label{fig:The-request-latencies}
\end{figure}

\paragraph{Average Latency} Furthermore, our framework calculates the average time taken to process GPU requests from issuance to completion. As depicted in Figure \ref{fig:The-request-latencies}, this metric encompasses both the waiting time and the execution time. The latter refers to the duration that a GPU engine requires to execute the request. To calculate the average latency, denoted by L in the formula below, our framework sums the waiting (W) and execution (E) times for all requests processed within a given observation period. This sum is then divided by the number of requests, n, to yield the average latency value.
\begin{equation*}
L=\sum_{i}\frac{W_{i}+E_{i}}{n}
\end{equation*}
\vspace{-15pt}

\paragraph{GPU Memory Usage} 
The i915 GPU driver provides deep insights into memory usage through specific events (e.g., i915\_gem\_object\_create and i915\_gem\_object\_destroy). By tracking these events, we can measure the total memory allocated and subsequently released by Graphics Execution Manager (GEM) objects, representing the dynamic footprint of GPU memory usage. This metric is crucial for understanding how memory allocation and usage impact performance, particularly in virtualized environments where optimizing resource utilization is paramount. Using this metric helps pinpoint potential memory-related bottlenecks and inefficiencies both in the host machine and within VMs.

\subsection{Visualization \label{visualisation-Subsystem}}
The presentation of information in a trace analysis framework should be made in a manner that enhances the user's comprehension of the analyzed system. Hence, our framework provides various views, each tailored to illustrate a specific aspect of GVT-g operations. These views have been integrated as plug-ins within Trace Compass, a versatile open-source trace visualization tool. Examples of the visualizations our framework provides include the \textit{GPU usage view}, an XY chart that graphically represents the percentage of utilization of vGPUs and pGPUs. Another visualization is the \textit{latency view}, which displays the durations spent by GPU requests in queues and during execution. This view employs a time graph format, where colored rectangles represent the latencies of individual GPU requests, each marked with its corresponding request ID. Our framework can display the transitions of a request between queues using arrows from the source to the destination.


\section{Use Cases}\label{Use-Cases}
This section aims to demonstrate the practicality and effectiveness of our framework in diagnosing and examining performance issues in GPU-accelerated VMs. We detail the configurations used in our investigations in Table \ref{tab:experimental-configurations}.

\begin{table}[H]
\caption{ Experimental configurations of the hardware and software.}
\label{tab:experimental-configurations}
\centering
\small
\begin{tabular*}{\linewidth}{@{\extracolsep\fill}llll@{\extracolsep\fill}}
\toprule
\multicolumn{2}{l}{\textbf{Host Machine}} & \multicolumn{2}{l}{\textbf{Virtual Machine}} \\
\midrule
~~~CPU & Intel Core i7-7500U @2.70GHz  & vCPU & 2 \\
~~~GPU & HD Graphics 620 & vGPU Mem. & Low: 64 MB/High: 384 MB   \\
~~~RAM & 16 GB  & RAM & 4 GB   \\
~~~OS & Ubuntu 16.04 (Kernel 4.14.15) & OS & Ubuntu 16.04 (Kernel 4.15)   \\
~~~Qemu & v2.11.1 \\
~~~LTTng & v2.10.8 \\
\bottomrule
\end{tabular*}
\end{table}

\subsection{GVT-g Scheduling}
\begin{figure}[H]
\includegraphics[width=13cm]{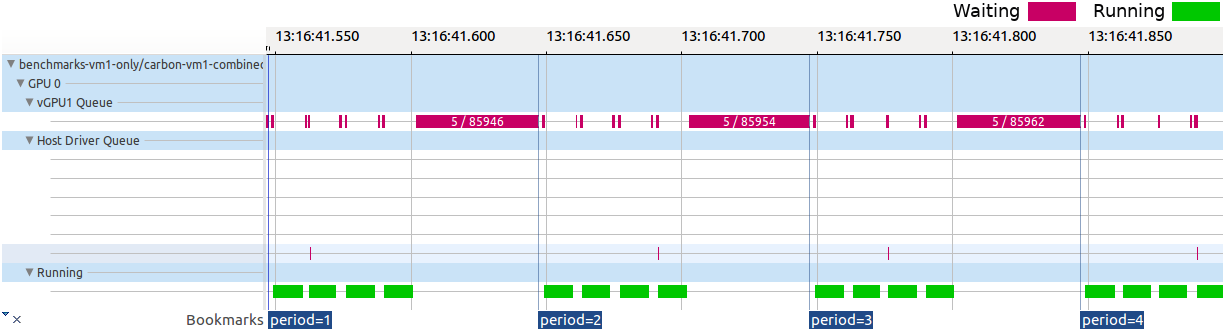}
\caption{Time-graph view illustrating the wait times of GPU requests in GVT-g and host driver queues, as well as their respective execution times. It also shows how the scheduler defers the execution of vGPU1 requests to the next period when its allocated time slice is exhausted.}
\label{fig:requests-control-flow}
\end{figure}

In this case study, we use our framework to examine the internal mechanics of the GVT-g scheduler, as its operations may impact the performance of GPU virtualization solutions. Specifically, GVT-g uses a kernel thread named \textit{gvt\_service\_thr} and multiple additional threads, which are referred to as \textit{workloads thread X}, where ``X'' represents a specific engine supported by the pGPU, such as render or video command streamers. The \textit{gvt\_service\_thr} is responsible for managing vGPU scheduling and context switching, and the \textit{workloads thread X} threads are initiated when a vGPU is activated. These threads concurrently monitor the waiting queue, facilitating the transfer of GPU requests to the host GPU driver, as depicted in Figure \ref{fig:requests-control-flow}.

\begin{figure}[H]
\includegraphics[width=8.5cm]{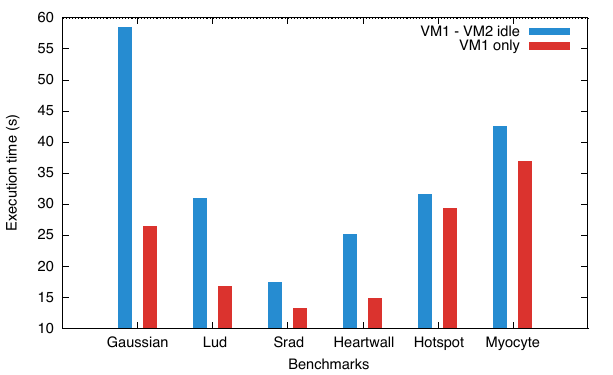}
\caption{Latencies of benchmarks when executed in multi-VM and single-VM contexts.}
\label{fig:latencies-multi-vm-only-vm}
\end{figure}

Our initial experiment involved two GPU-accelerated virtual machines (VM1 and VM2) sharing a single pGPU via GVT-g, with each VM hosting a similarly configured vGPU and equal resource allocation weights. These weights, assigned by GVT-g at instantiation, determine each vGPU's share of pGPU resources. We executed a series of benchmarks from the Rodinia GPU benchmark suite \cite{Che2009} on VM1. We repeated these benchmarks after completely shutting down VM2. Figure 
 \ref{fig:latencies-multi-vm-only-vm} presents histograms comparing the benchmarks' execution times in both scenarios. Notably, the benchmarks executed slower when VM2 was idle than when it was shut down, with significant time discrepancies in the Gaussian, Lud, and Heartwall benchmarks. This study considered a VM idle if it did not engage in significant GPU workloads.

\begin{figure}[H]
\includegraphics[width=13cm]{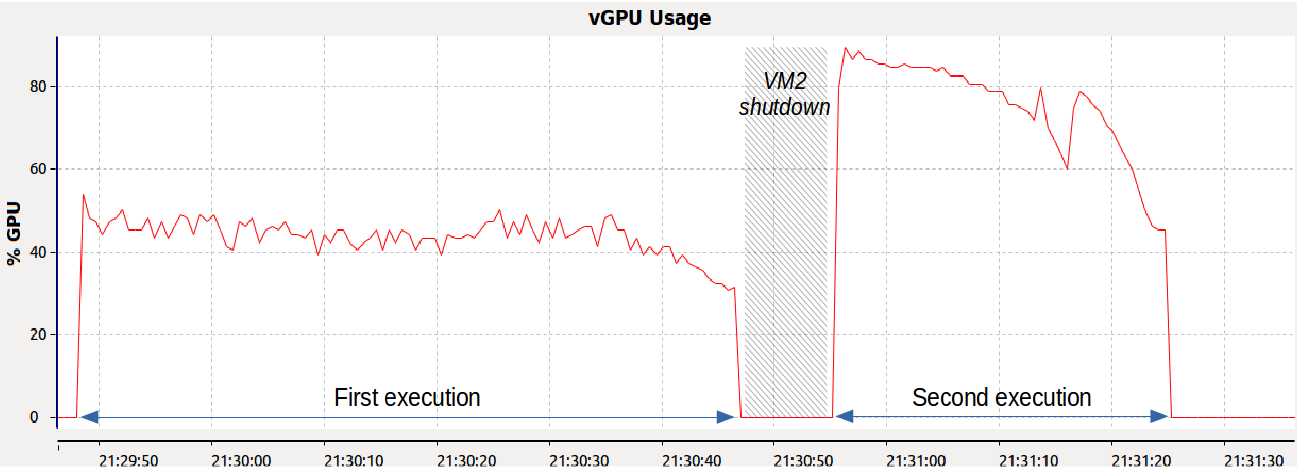}
\caption{XY view depicting the GPU usage percentages for two consecutive runs of the Gaussian benchmark in VM1. It also highlights the significant increase in GPU usage following the shutdown of VM2, which led to faster execution of the benchmark.}
\label{fig:Gpu-usage-gaussian}
\end{figure}

To investigate the performance difference, we analyzed the execution of the Gaussian benchmark on VM1 using our framework. Figure \ref{fig:Gpu-usage-gaussian} illustrates the benchmark's execution duration and GPU utilization. Initially, the benchmark lasted 59 seconds with an average GPU utilization of about 45\%. In contrast, after shutting down VM2, the benchmark was completed in 27 seconds with a GPU utilization near 80\%. This comparison highlights the performance impact of VM2's concurrent operation, despite it being inactive.

Further examination of the GVT-g scheduler's operations revealed its reliance on time multiplexing for fair pGPU resource distribution among vGPUs. The scheduler, which was operated via the \textit{gvt\_service\_thr} thread, simultaneously managed all of the GPU engines to minimize scheduling complexity and avoid synchronization bottlenecks. It assigned time slices to vGPUs based on their weights, deducting the aggregate execution time of a vGPU's requests from its time slice (see the formula below). If a vGPU depletes its time slice within a certain interval (100 ms), the scheduler suspends its operations, recalibrating time slices every ten intervals (one second) and ignoring past debts or credits.
\vspace{-9pt}

\vspace{-8pt}

\vspace{-12pt}
\begin{equation*} \label{form-time-slice}
 Time\: Slice = 100~\text{ms} \times \frac{vGPU_{weight}}{\sum_{i}^{} GPU^i_{weight}}
\end{equation*}

\begin{Example}
Let us consider two vGPUs sharing the host GPU's resources, and each is assigned an identical weight. In this scenario, the GVT-g scheduler assigns each vGPU a 50 ms time slice per period. If one vGPU is removed, the remaining one receives the entire 100 ms time slice per period, fully utilizing the available resources.
\end{Example}

The GVT-g scheduler follows a round-robin policy, cycling every 1 millisecond to inspect vGPU queues for new requests. Upon finding requests, the scheduler activates the underlying vGPU for processing if its time slice remains. Concurrently, the `workloads threads X' handle the active vGPU's requests and update statistics for the previously active vGPU. If a vGPU exhausts its time slice, the scheduler then schedules another vGPU, as shown in Figure \ref{fig:requests-control-flow}.

In short, our analysis shows that a VM cannot use the idle time slice of another VM due to the GVT-g scheduler policy. While this aligns with typical cloud computing models, it is suboptimal for private cloud data centers seeking to maximize vGPU acceleration. We propose enhancing the GVT-g scheduler with an algorithm that allows a VM to process its GPU workloads when the host GPU resources are unoccupied. A configuration option can be set during startup to allow GVT-g to apply the appropriate scheduling algorithm based on the deployment environment, enabling more efficient GPU resource management.

\subsection{Analysis of Contention for Resources }
\begin{figure}[H]
\includegraphics[width=8cm]{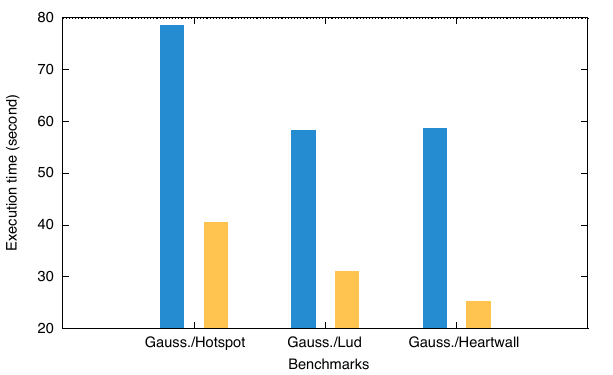}
\caption{The execution time of the Gaussian benchmark (VM1) increased when it was concurrently executed with the Hotspot benchmark (VM2). Contention for GPU resources between the two VMs is a plausible cause.}
\label{fig:simultaneous-executions-benchmarks}
\end{figure}

\begin{figure}[H]
\includegraphics[width=13cm]{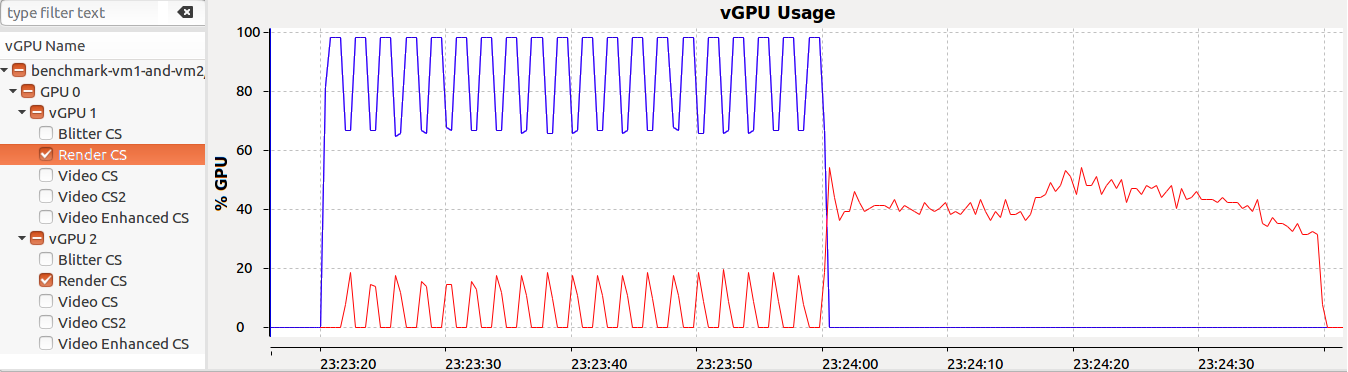}
\caption{View depicting the GPU usage during concurrent runs of the Gaussian/VM1 (red) and Hotspot/VM2 (blue) benchmarks. It also demonstrates the negative impact of the size of workloads executed by vGPU2 on the performance of vGPU1.}
\label{fig:GPU-usage-Gaussian-Hotspot}
\end{figure}

This second use case looks into the extent to which the performance of a vGPU-enabled VM can be affected by the presence of other VMs. In our experiment, two collocated VMs shared host GPU resources via GVT-g, and their vGPUs were configured identically, as detailed in Table \ref{tab:experimental-configurations}. On VM1, we ran the Gaussian benchmark, while VM2 executed a variety of benchmarks characterized by various intensities of host GPU resource usage (e.g., varying workload sizes, diverse request frequencies, etc.). For example, Figure \ref{fig:simultaneous-executions-benchmarks} demonstrates the effects of workload sizes from these benchmarks on the Gaussian benchmark's execution time. Notably, the execution time of the Gaussian benchmark increased significantly when run concurrently with the Hotspot benchmark. Our framework view, which is depicted in Figure~\ref{fig:GPU-usage-Gaussian-Hotspot}, further illustrates that during the Hotspot benchmark's execution, GPU usage for the Gaussian benchmark did not exceed 20\%, but this rose to 40\% after the Hotspot \mbox{benchmark ended.}
\vspace{-5pt}

\vspace{-6pt}

This experiment shows that the performance of one vGPU can be considerably affected by another vGPU that is co-located. Our analysis indicates that this is primarily caused by the disparity in workload sizes: approximately 18 ms for vGPU1 (Gaussian benchmark) versus 800 ms for vGPU2 (Hotspot benchmark). The i915 GPU driver's scheduler lacks preemptive multitasking capabilities, meaning that it cannot interrupt an ongoing workload. Thus, the GVT-g scheduler uses a coarse-grained sharing approach for the host GPU time to compensate. Time slices are allocated to vGPUs in proportion to their weights, and the execution times of GPU requests are subtracted from these slices. Outstanding balances (debts and lefts) from previous periods are carried over for up to ten periods, explaining why the large workloads from vGPU2 adversely affected vGPU1's performance.

Despite enhancing scheduling fairness among vGPUs and, thus, improving VM application responsiveness, this method's drawback lies in its coarse-grained approach to host GPU time-sharing. This becomes particularly problematic when a vGPU's workload size exceeds the period length. 

In summary, this use case demonstrates that the GVT-g scheduler struggles to maintain fairness between vGPUs under certain conditions, which is largely due to the i915 driver's lack of preemption support and the implementation of its coarse-grained sharing mechanism. Because the scheduling algorithm's focus is to maintain interactive program responsiveness, we suggest disabling the debt and leftover management feature when VMs are tasked with running non-interactive tasks.

\subsection{Virtual Machine Placement}
Live migration, a key feature supported by most modern hypervisors, involves relocating VMs to balance the load on physical host machines. This process not only optimizes physical resource utilization within data centers but also ensures adherence to service-level agreements. VM placement, which is a critical component of VM migration, entails selecting the most suitable physical machine to host a VM. Many placement algorithms have been developed over time to maximize resource utilization in cloud data centers and minimize energy consumption. The efficiency of a placement algorithm is paramount, as studies show a strong correlation between resource optimization and the reduction of operational costs in data centers
\cite{Masdari2016}.

Effective VM placement algorithms often depend on sophisticated monitoring tools. These tools assess VM resource requirements and identify performance issues stemming from resource overcommitment. Overcommitment leads to VMs competing for shared resources, causing frequent interruptions and prolonged execution times for running programs. To mitigate this, several algorithms co-locate VMs with complementary resource demands on the same host. For example, pairing a CPU-intensive VM with a disk or network-intensive VM can optimize resource utilization and enhance overall performance. Similarly, aligning VMs that alternate between CPU and GPU resource demands within the same host proves beneficial. Hence, co-locating CPU- and GPU-bound VMs optimizes the use of both processing units, which lowers the frequency of VM preemption and boosts the overall system performance.

\begin{figure}[H]
\includegraphics[width=13cm]{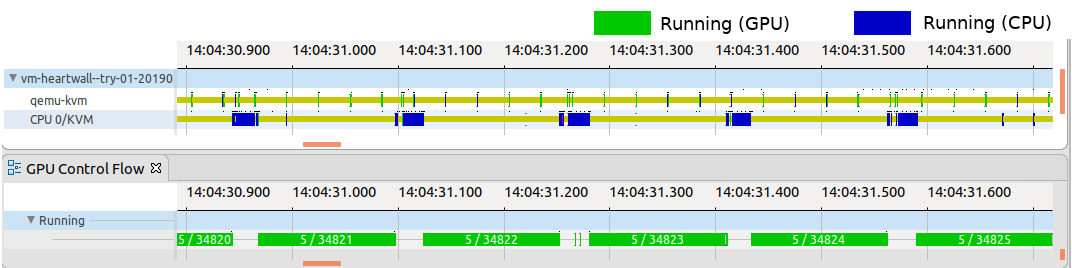}
\caption{View showing a VM deployed in our private cloud that alternated between CPU-intensive and GPU-accelerated computations.}

\label{fig:gpu-cpu-exec-alternance}
\end{figure}

Seeking to optimize VM placement within our private cloud, which was managed via OpenStack~\cite{OpenStack2024}, we leveraged our performance analysis framework to incorporate GPU usage data from running VMs as a key input parameter. This optimization involved migrating VMs that exceeded a predefined CPU usage threshold over a certain period to host machines that were mostly engaged with GPU-intensive tasks. This reallocation significantly enhanced the resource utilization across our cloud infrastructure and reduced the execution times of applications within the VMs. The view from our tool in Figure \ref{fig:gpu-cpu-exec-alternance} illustrates this strategy, showcasing a VM that alternates between CPU and GPU computing phases.

\begin{figure}[H]
\includegraphics[width=13cm]{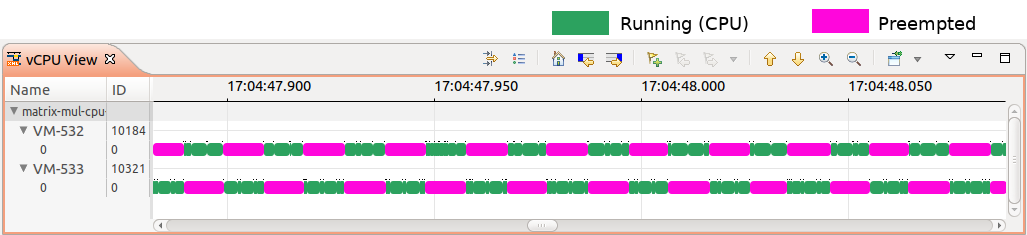}
\caption{View showing how VM-532 and VM-533  preempted each other as they competed for the same CPU resources.}
\label{fig:vms-cpu-contention}
\end{figure}

\begin{figure}[H]
\includegraphics[width=13cm]{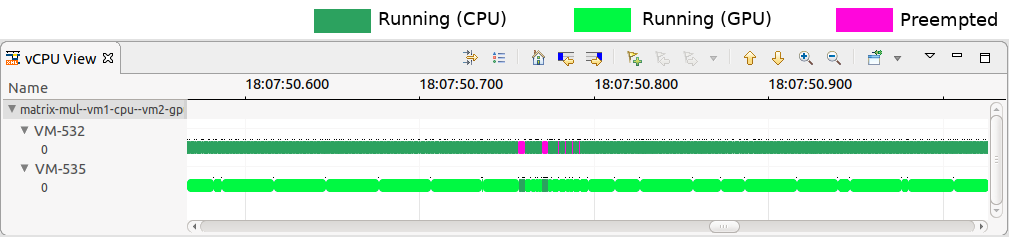}
\caption{Our analysis shows that after applying the optimized VM placement algorithm, VM-532 was rarely preempted by the GPU-intensive VM-535.}
\label{fig:vms-cpu-gpu-preemption}
\end{figure}

To validate our approach, we used the \textit{sched\_switch}, \textit{kvm\_entry}, and \textit{kvm\_exit} events to compute the number of VM preemptions. Our method involved tracing the operating system kernels of the cloud host machines and categorizing VMs as either GPU-intensive or CPU-intensive using our analysis framework. For example, we observed several CPU-intensive VMs, such as VM-532 and VM-533, co-located on the same host and frequently preempting each other for CPU resources, as depicted in Figure \ref{fig:vms-cpu-contention}. Conversely, other hosts  housing VMs primarily running GPU-intensive programs with minimal CPU dependency were identified. We then tagged VMs accordingly as 'CPU-intensive' and 'GPU-intensive'. Our refined placement algorithm reassigned CPU-intensive VMs to hosts with GPU-intensive VMs when CPU-overcommitment-induced preemptions surpassed a specific threshold. After the implementation of this optimized algorithm, a marked decrease in VM preemptions was observed. Figure \ref{fig:vms-cpu-gpu-preemption} illustrates this improvement, showing a significant reduction in the preemptions of VM-532 after VM-535's relocation, with preemption counts dropping from 23,125 to 15,087.
\vspace{-8pt}

\vspace{-15pt}

\vspace{-15pt}

\section{Tracing Cost Analysis}\label{Tracing-Cost-Analysis}
This section examines the cost implications associated with the use of our framework---particularly the trace data collection phase. As noted before, our approach relies mainly on efficient host kernel tracing. However, the overhead introduced by this tracing must be minimal to prevent any significant impact on the performance or behavior of the system under observation.

To estimate the overhead incurred by our tracing method, we conducted a series of experiments using a VM, the specifications of which are detailed in Table \ref{tab:experimental-configurations}. This VM was utilized to execute selected benchmarks from the Rodinia benchmark suite \cite{Che2009}, which were chosen due to their varied workload sizes and submission frequencies. Our results, as summarized in Table \ref{tab:tracing-cost}, indicate that the operational overhead from active tracepoints was modest, averaging around 1.01\%. This low overhead confirmed the efficiency of our tracing approach in gathering runtime data without significantly burdening the system.

Storing trace files can be problematic for most tracing tools, as they generate data in large amounts. This storage cost is mostly influenced  by two factors: the event frequency and the chosen storage format. Our findings, which are illustrated in Table \ref{tab:tracing-cost}, show that the size of the trace files did not directly correlate with the benchmark execution times. Rather, it was more closely related to the number of events generated, which, in our context, corresponded to the volume of GPU request submissions. Owing to our strategy of selectively activating a limited number of tracepoints, we managed to keep the trace file sizes within reasonable limits, thus mitigating potential storage issues.

\begin{table}[H]
\tablesize{\small} 
\caption{Tracing overhead.\label{tab:tracing-cost}}
		\begin{tabularx}{\linewidth}{CCCCCC}
			\toprule
			\textbf{Benchmark}	& \textbf{Exec Time (ms)}	& \textbf{Exec Time w/ Tracing 
 (ms)} & \textbf{\# 
 Events} & \textbf{Trace Size (KB)} & \textbf{Overhead* (\%)}\\
			\midrule
			Gaussian 
    & 26,
393 & 26,906 & 49,255 & 1228 & 1.906 \\
			Hotspot     & 29,455 & 30,161 & 9122  & 248  & 2.341 \\
			Lud         & 16,736 & 16,862 & 19,866 & 512  & 0.743 \\
			Myocyte     & 36,935 & 37,061 & 158,587& 3788 & 0.341 \\
			Srad        & 13,206 & 13,238 & 73,973 & 1843 & 0.237 \\
			Heartwall   & 14,795 & 14,873 & 4110  & 148  & 0.525 \\ 
			\bottomrule
		\end{tabularx}
	\noindent{\footnotesize{* 
 The overhead is computed as the percentage increase in execution time with tracing enabled.}}
\end{table}
\vspace{-16pt}

\section{Conclusions}\label{Conclusion}
A GPU is a powerful computing unit with remarkable characteristics, including a highly parallel architecture and energy efficiency. Hence, this device has found extensive application in diverse domains, such as heterogeneous and embedded systems, where it greatly improves application performance. These applications range from autonomous driving to media transcoding and graphics-intensive applications. The interesting attributes of this accelerator have prompted the introduction of vGPU-accelerated VMs, opening up opportunities to address CPU bottlenecks in data centers and enabling end-users to harness the power of high-performance computing devices for a variety of applications.

While our review of the existing literature shows various performance analysis tools for GPUs, support for vGPUs remains a significant gap. Unfortunately, the only tool offering vGPU support, NVIDIA Nsight Systems, is closed-source software and lacks publicly available documentation. To address this gap in research, our study introduces a novel diagnostic tool that is specifically designed to analyze vGPU performance. Utilizing GVT-g, Intel's virtualization technology for on-die GPUs, as a foundational platform, our framework provides detailed insights into GVT-g operations. It facilitates an exhaustive analysis of VMs equipped with vGPUs, thereby enhancing the understanding and optimization of their performance.

Our approach leverages LTTng, a low-overhead tracer, to collect the low-level data required for our analyses. Through this tool, we traced the host machine kernel to gain a better understanding of vGPU resource utilization. Furthermore, we  developed a suite of auto-synchronized views in Trace Compass to unveil the intricate dynamics of vGPU-accelerated VMs as they interact with the host GPU resources. Finally, our analysis reveals that the tracing overhead is minimal  at approximately 1.01\%. Looking ahead, future iterations of our work will expand this approach to encompass other GPU virtualization technologies, such as MxGPU and NVIDIA Grid. These technologies have gained prominence in the data centers of major cloud providers, making our research even more pertinent and valuable in the evolving landscape of GPU virtualization.

\vspace{6pt} 




\authorcontributions{Conceptualization, A.B.; Methodology, A.B.; Software, A.B.; Validation, A.B.; Investigation, A.B.; Writing – original draft, A.B.; Supervision, M.D.; Funding acquisition, M.D. All authors have read and agreed to the published version of the manuscript. 
}

\funding{
This research was financed by the Natural Sciences and Engineering Research Council of Canada (NSERC), through the NSERC Alliance project 554158-20, in partnership with Prompt, Ericsson, Ciena, AMD, and EfficiOS. The APC was funded by Adel Belkhiri.}

\dataavailability{The source codes of the proposed framework are available on the author’s {GitHub:} 
 \url{https://github.com/adel-belkhiri} (accessed on 20 February 2024).}

\acknowledgments{We would like to thank the Natural Sciences and Engineering Research Council of Canada (NSERC), Prompt, Ericsson, Ciena, AMD, and EfficiOS for supporting this research.}


\conflictsofinterest{The authors declare no conflicts of interest.} 




\begin{adjustwidth}{-\extralength}{0cm}
\reftitle{References}

\PublishersNote{}
\end{adjustwidth}
\end{document}